\documentclass[12pt]{article}
\usepackage{amssymb}
\usepackage{latexsym}
\def\a{\alpha}
\def\b{\beta}
\def\d{\delta}
\def\ep{\varepsilon}
\def\f{\varphi}
\def\e{\eta}
\def\l{\lambda}
\def\L{\Lambda}
\def\p{\psi}
\def\r{\rho}
\def\s{\sigma}
\def\S{\Sigma}
\def\x{\xi}
\def\N{\mathbb{N}}
\def\Z{\mathbb{Z}}

\def\R{\mathbb{R}}

\newcommand{\MA}{{\mathcal{A}}}
\newcommand{\MI}{{\mathcal{I}}}
\newcommand{\MP}{{\mathcal{P}}}
\newcommand{\MR}{{\mathcal{R}}}
\newcommand{\MS}{{\mathcal{S}}}

\newcommand{\card}{\mbox{card}}

\newtheorem{defn}{{\sc Definition}}[section]
\newtheorem{thm}{{\sc Theorem}}[section]
\newtheorem{lem}[thm]{{\sc Lemma}}
\newtheorem{pro}[thm]{{\sc Proposition}}
\newtheorem{cor}[thm]{{\sc Corollary}}
\newtheorem{rem}{\sc Remark}[section]
\newtheorem{eg}{\sc Example}[section]

\def\proof{{\noindent\em Proof.} }
\def\qed{\hfill $\Box$  \newline \vspace{2mm}}

\catcode`\@=11
\@addtoreset{equation}{section}
\begin{document}

\title{ Symbolic Representations  of Iterated Maps }

\author{}
\date{November 14,  2000 }

\maketitle


\begin{center}

{\large Xin-Chu Fu$^{1}$}, {\large Weiping Lu$^{2}$}, 
{\large Peter Ashwin$^{1}$} and {\large Jinqiao Duan$^{3}$} 

\vspace{0.2cm}
  1. {\em   School of Mathematical Sciences, Laver Building\\
       University of Exeter, Exeter EX4 4QJ, UK}\\

  2. {\em  Department of Physics, Heriot-Watt University \\
              Riccarton, Edinburgh EH14 4AS, UK} \\
  
  3. {\em  Department of Applied Mathematics\\
        Illinois Institute of Technology, Chicago, IL 60616, USA}
\vspace{0.2cm}              

\end{center}

\begin{abstract}
\noindent  This paper presents a general and systematic discussion of 
various symbolic representations of iterated maps through subshifts.
We give a unified model for all continuous maps on a metric space, 
by representing 
a map through a general subshift over usually an uncountable alphabet. 
It is shown that at most the second order representation
is enough for a continuous map. 
In particular, it is shown that the dynamics of one-dimensional 
continuous maps to a great extent can be transformed to 
the study of subshift structure of a general symbolic 
dynamics system. By introducing distillations, partial representations 
of some general continuous maps are obtained. 
Finally, partitions and representations of a class of discontinuous maps,
 piecewise continuous maps are discussed, and as examples, a representation 
 of the Gauss map via a full shift over a countable alphabet and 
 representations of interval exchange transformations as subshifts of infinite 
 type are given.

\vspace{0.3cm}

\noindent {\bf Key words:} \  Continuous Map, Discontinuous Map, 
                Representation, Subshift, Symbolic Dynamics  

\end{abstract}

\section{Introduction}

It has long been known that interesting 
 behaviour can occur when iterating   continuous maps.
Such maps define discrete dynamical systems, which 
have been used as simplified prototypical
models for some engineering and biological
processes; see, e.g.~\cite{May, Hao89, Fu98}.
Through the Poincar\'e section maps, discrete dynamical systems have also 
been used to study continuous dynamical systems, e.g.~\cite{GH83, Wiggins88}.

Considerable progress has been made in the last two decades in the  
understanding of dynamical behaviour of nonlinear continuous maps. 
These systems, while mostly studied individually because of their 
distinct nature of nonlinear interactions, show similar dynamical features, 
especially when the parameters of the systems are close to some critical values 
where abrupt change in behavior takes place. The universality of such behavior 
 has been a key subject in the study of nonlinear dynamics. A mathematical 
framework has, however, yet to be established under which a class of dynamical 
systems, such as one-dimensional continuous maps, can be described by a unified 
model. Such a framework will improve our understanding of general properties
of dynamical systems and may be useful in our effort to classify
dynamical systems.

Shifts and subshifts defined on a space of abstract symbols
are special discrete dynamical systems which are called
symbolic dynamics systems. Symbolic dynamics is a powerful tool to study 
more general discrete dynamical systems, because the latter often contain 
invariant subsets on which the dynamics is similar or even 
equivalent to a shift or subshift.  
Moreover, there are a number of definitions of chaos, namely
(1) the Li-Yorke definition;
(2) Devaney's definition; (3) topological mixing;
(4) Smale's horseshoe; (5) transversal homoclinic points;  and 
(6) symbolic dynamics. 
The symbolic dynamics definition is especially important of these definitions, 
as it unifies aspects of many of the definitions. More precisely, it 
implies the first three definitions, is topologically conjugate to
the fourth one and occurs as a subsystem of the fifth. 
Furthermore (see for example Ford \cite{Alekseev})
symbolic dynamics is very important for analysis of applications of 
nonlinear dynamics in physical sciences.

For a dynamical system, we can study it either directly or via other 
systems which are better understood. Symbolic representations are methods to 
study dynamical systems through shifts and subshifts.
In the study of hyperbolic dynamics of homeomorphisms, symbolic dynamics is 
one of the most fundamental models. Since the discovery of the Markov 
partitions of the two dimensional torus by Berg \cite{Berg} and the  
related work by Adler and Weiss \cite{Adler2}, symbolic representations of  
hyperbolic systems through the Markov partitions have been studied 
extensively (e.g., see \cite{Adler}). 
 
The equivalence between Smale's horseshoe and the symbolic dynamical system 
$(\S(2), \s)$ (see \cite{Smale}) implies that the former has a symbolic 
representation through the full shift $(\S(2), \s)$. A similar symbolic 
representation has also been revealed by Wiggins (see \cite{Wiggins88}) 
on higher dimensional versions of the Smale's horseshoe. Earlier works 
in \cite{Zhang} and \cite{Fu96} 
established that, under certain conditions, the 
restrictions of a general continuous self-map $f: X \rightarrow X$ to some 
horseshoe-like invariant subsets are topologically conjugate to 
$\s|_{\S(N)}$ or $\s|_{\S(\Z_+)}$ 
(see \cite{Fu96}), where $(\S(\Z_+), \s)$ is the symbolic dynamics  system 
with a countable  alphabet. These results actually demonstrated the partial 
symbolic representation of a class of maps as a full shift over a 
countable alphabet.

Motivated by the above work, we study in this paper symbolic representations of 
continuous self-maps 
by using a more general and systematic approach. We present a unified model for 
all  continuous maps on a metric space, by representing a map as a general 
subshift $(\S(X),\s)$. We show that the subshifts of $(\S(X), \s)$ may be 
used as such a unified model for all continuous self-maps on a metric space $X$. 
And it is shown that at most the second order representation
is enough for a continuous map.
In particular, when $X$ is a closed interval, we show that the dynamics of 
one-dimensional continuous maps to a great extent can be transformed to
the study of subshifts of a symbolic dynamical system.
We also discuss quasi-representations, and by introducing distillations, 
partial representations of some general continuous maps are obtained. 
Finally, partitions and representations of a class of discontinuous maps,
 piecewise continuous maps, are discussed, and as examples, a regular 
 representation of the Gauss map via a full shift over a countable alphabet 
 and regular representations of interval exchange transformations as 
 subshifts of infinite type are given.

\section{Some Definitions, Notation and Lemmas}

Before turning to the next section to discuss representation theorems, we 
recall some definitions, introduce some notation, and provide some lemmas. 

Let $(X,d)$ be a metric 
space and denote by $\S(X)$ the space $X^{\Z_+}$ which
consists of functions from the nonnegative integers $\Z_+$ to the
metric space $X$. $x \in \S(X)$ may thus be denoted by $x=(x_0, x_1, \cdots, 
x_i, \cdots)$, $x_i \in X, i \ge 0$. Further, let $\S(X)$ be endowed with the 
product
topology, so $\S(X)$ is metrizable. The metric on $\S(X)$ can
be chosen to be
$$ 
\r(x, y)=\sum^{+\infty}_{i=0} \frac{1}{2^i} \frac{d(x_i, y_i)}{1+
d(x_i, y_i)}, \;\;\;\; x=(x_0, x_1, \cdots), \;\; y=(y_0,y_1, \cdots) \in
\S(X).
$$

The shift map $\s: \S(X) \rightarrow  \S(X)$ is defined by
$(\s(x))_i=x_{i+1}, \; i=0,1, \cdots.$
Since $ \r (\s(x), \s(y)) \le 2 \r(x, y), $
$\s$ is continuous. $(\S(X), \s)$ is a general symbolic dynamics system 
(see \cite{Wiggins88, Fu92, Fu98}). We call $X$ the 
symbol space or alphabet, and $\S(X)$ the symbol sequence space.

When $X$ is chosen as $\{0,1,\cdots, N-1 \}$ and the metric $d$ on
$\{0,1,\cdots, N-1 \}$ is the discrete metric:
$$ d(m,n)=\left\{\begin{array}{ll} 0, \;\;\;\; m=n\\ 1, \;\;\;\; m\ne n
 \end{array} \right., $$
then $(\S(X), \s)$ becomes the usual symbolic dynamics system
$(\S(N), \s)$ as in \cite{GH83}.

Let $\S \subseteq \S(X)$ be closed, and invariant for $\s$, i.e., $\s(\S) 
\subseteq \S $, then $(\S, \s)$ forms a subsystem of $(\S(X), \s)$. We call 
$(\S, \s)$ a subshift of the full shift $(\S(X), \s)$, denoted by $(\S, \s) \le 
(\S(X), \s) $.

We denote by $C(X)$ the set of all continuous self-maps on $X$ and $M(X)$ 
the set of all self-maps on $X$. We also call the iteration system of an 
$f \in M(X)$ a dynamical system, denoted by $(X, f)$.
For two dynamical systems $f_1 : X_1 \rightarrow X_1 $ and $f_2 : X_2 
\rightarrow X_2 $, where $X_i$ are metric spaces and $f_i \in M(X_i), i=1,2,$ 
if there exists a homeomorphism $h: X_1 \rightarrow X_2 $ such that $hf_1=f_2h$, 
then we call $f_1$ is topologically conjugate to $f_2$, denoted by $f_1 \sim 
f_2$. If $h$ is continuous and surjective, but not necessarily invertible, 
and if $hf_1=f_2h$, 
then we call $f_1$ is topologically semi-conjugate to $f_2$, and also call 
$f_2$ a factor of $f_1$ (and $f_1$ an extension of $f_2$). 
We call $f_1$ and $f_2$ are weakly conjugate if each is a factor of the other.

We give the following definitions for various symbolic representations:

\begin{defn}
If for an $f \in M(X)$, there exists a subshift $(\S, \s)$ of certain a
symbolic dynamical system and a surjective map $h: \S \to X$, and a countable 
subset $D \subseteq \S$, such that $h$ is one-to-one and continuous on 
$\S\setminus D$, finite-to-one on $D$, and 
$f\circ h = h \circ \s $, then we call the subshift $(\S, \s)$ a 
{\rm regular representation} of the dynamical system $(X, f)$.
If $D=\varnothing$, we call the subshift $(\S, \s)$ a
{\rm faithful representation} of the dynamical system $(X, f)$.

If $h$ is a topological conjugacy, we call $(\S, \s)$ a 
{\rm conjugate representation} of $(X, f)$.

If $\s|_{\S}$ and $f$ are weakly conjugate, we call $(\S, \s)$ a 
{\rm weakly conjugate representation} of $(X, f)$.

If $h$ is a topological semi-conjugacy, i.e., $(X, f)$ is a factor of 
$(\S, \s)$, we call $(\S, \s)$ a {\rm semi-conjugate representation} of 
$(X, f)$.

If $(\S, \s)$ is a factor of $(X, f)$, we call $(\S, \s)$ a 
{\rm quasi-representation} of $(X, f)$.

In all the above representations, we call a correspondence between 
symbol sequences in $\S$ and points in $X$ (either from $\S$ to $X$ or from 
$X$ to $\S$) a {\rm coding}.
\end{defn}

{}From the definition above, among the six representations, conjugate 
representation is the strongest one, since it implies all the others. A weakly 
conjugate representation is always a semi-conjugate representation and a 
quasi-representation. A faithful representation is also a semi-conjugate 
representation. Quasi-representation is the weakest one, but it still tells us 
some information. For example, if $(\S, \s)$ is a quasi-representation of 
$(X, f)$, then $h_{\rm top} (f)\ge h_{\rm top} (\s |_{\S})$, while the 
latter is usually easier to calculate 
($h_{\rm top} (\cdot)$ denotes the topological entropy).

In the definition of a regular representation, we allow $h$ to 
be possibly discountinuous 
on $D$ so that we can apply this definition to symbolic representations of some 
discontinuous maps.

Note that weak conjugacy is strictly weaker than conjugacy (\cite{Lind}). 
For example, the Fibonacci subshift (consists of all sequences of $0$'s and 
$1$'s with $1$'s separated by $0$'s) and the even subshift 
(consists of all sequences of $0$'s and $1$'s with $1$'s separated by an 
even number of $0$'s) are weakly conjugate, but they are not conjugate since 
one is a subshift of finite type and the other is a subshift of infinite 
type.

Throughout this paper, for $a \in I=[0, 1]$, we denote by $r_m(a)$ the $m$-adic 
fraction
$$
r_m(a)=\sum^{+\infty}_{i=0} \frac{a_i}{m^{i+1}} \ ,
$$ 
and when $a\ne 0$, we require that there are infinite number of non-zero 
$a_i$'s, $ i \ge 0$, and denote $a_i$ by $r^i_m(a)$.  Unless indicated 
otherwise, we will always make this choice when we write an $m$-adic fraction 
expansion of a real number.

For an integer sequence $x \in \S(N), 
x=(x_0, x_1, \cdots, x_i, \cdots)$, we also denote in form by $r_m(x)$ the 
$m$-adic  fraction
$$
r_m(x)=\sum^{+\infty}_{i=0} \frac{x_i}{m^{i+1}} \ , ~~\mbox{where}  \  m \ge N.
$$

For $ a, b \in I,$ let 
\begin{eqnarray*}
d_0(a, b) & = & |a-b|, \\
d_1(a, b) & = & \sum^{+\infty}_{i=0}\frac{1}{2^i}|r^i_2(a)-r^i_2(b)|,
\end{eqnarray*} 
and denote by $I_k$ the metric space $(I, d_k),\ k=0,1.$ 

Define the metric $\r_k$ on $\S(I_k)$ as
$$
\r_k(x,y)=\sum^{+\infty}_{i=0} \frac{1}{2^i} 
\frac{d_k(x_i,y_i)}{1+d_k(x_i,y_i)}, \ \ k=0,1.
$$

The lemmas below are necessary for the proof 
of the theorems in the following sections.

\begin{lem}  \label{lem_metric}
Suppose $x=(x_0, x_1,\cdots,x_i, \cdots), 
y=(y_0, y_1,\cdots,y_i, \cdots) \in \S(X)$, then
\begin{eqnarray*}
\r(x,y) < \frac{1}{2^{2N+1}} ~~ & \Longrightarrow &~~  
d(x_i, y_i)< \frac{1}{2^N},~\forall ~i \le N;\\
\mbox{For}~ \d > 0, ~d(x_i, y_i)< \d,~\forall ~i \le N  ~~ & 
\Longrightarrow &~~  
\r(x,y) < 2 \d+ \frac{1}{2^N}.
\end{eqnarray*}
\end{lem} 
\proof
Otherwise, if there exists $i_0\le N,$ such that 
$d(x_{i_0}, y_{i_0})\ge 1/2^N, $ then 
$$
\r(x, y)\ge \frac{1}{2^{i_0}}\frac{\frac{1}{2^N}}{1+\frac{1}{2^N}}\ge 
\frac{1}{2^N}\frac{1}{1+2^N}\ge \frac{1}{2^{2N+1}} ~ .
$$
This is a contradiction. So we have the first inequalities.

We can check directly the second inequality.  \hfill    $\Box$

\begin{lem}  \label{lem_compact}
If $A\subseteq I$ is compact in $I_1$, then $A$ is also compact in $I_0$;
and if $\S\subseteq \S(I)$ is compact in $\S(I_1)$, 
then $\S$ is also compact in $\S(I_0)$.
\end{lem} 
\proof
$\forall a, b \in I$,
$$
d_0(a, b)=|a-b|=|\sum^{+\infty}_{i=0}2^{-(i+1)}(r^i_2(a)-r^i_2(b))|
\le \frac{1}{2}d_1(a,  b).
$$
So any limit points of $A$ in the metric $d_1$ are also limit points of $A$ 
in the metric $d_0$. Hence compactness of $A$ in $I_1$ implies compactness of 
$A$ in $I_0$.

Similarly, we can check the compactness of $\S$ in $\S(I_0)$.
\qed

We say a subshift $(\S, \s)$ satisfies the condition (\ref{condition}) if 
\begin{eqnarray}
\forall \ x=(x_0, x_1,\cdots ), \ y=(y_0,y_1, \cdots )\in \S, \ x=y \mbox{ if 
and only if } x_0 = y_0.     \label{condition}
\end{eqnarray}

\begin{lem}  \label{lem_condition}
For any $(\S, \s) \le (\S(I_1), \s)$, 
there exists $(\S^*, \s) \le (\S(I_1), \s)$ such that 
$\s|_{\S} \sim \s|_{\S^*}$, and $(\S^*, \s)$ satisfies 
the condition (\ref{condition}).
\end{lem} 
\proof
$\forall \x \in \S, \x=(\x_0, \x_1, \cdots, \x_i, \cdots). $ Rewrite $\x_i$ by
$$
\x_i=r_2(\x_i)=\sum^{+\infty}_{j=0} \frac{\x_{ij}}{2^{j+1}}, \ \ i=0, 1, \cdots.
$$
Let
\begin{eqnarray*}
r(\x)&=& 0.\x_{00}\x_{10}\x_{01}\cdots\x_{k0}\x_{k-1,1}\cdots\x_{1,k-1}\x_{0k}
\cdots   \ \ \  \mbox{(binary fraction),}\\
R(\x)&=& (r(\x), r(\s(\x)), \cdots,  r(\s^k(\x)), \cdots). 
\end{eqnarray*}
{}From $R(\x)=R(\e)$ we have $r(\x)=r(\e)$, thus $\x_{ij}=\e_{ij} \ , \       
  i,j=0,1, 
\cdots.$ As a result, $\x=\e$,  that is, $R$ is one-to-one. 

Whenever $\x, \e \in \S$ with $\r_1(\x, \e)<1/2^{2N+1}$, 
from Lemma~\ref{lem_metric} we have $d_1(\x_i, 
\e_i)<1/2^N, i\le N.$  So $\x_{ij}=\e_{ij} \ , \ i,j\le N$. Denote
\begin{eqnarray*}
r(\x)&=&\sum^{+\infty}_{k=0}2^{-(k+1)}(r(\x))_k \ , \\
r(\e)&=&\sum^{+\infty}_{k=0}2^{-(k+1)}(r(\e))_k \ ,
\end{eqnarray*}
we have
\begin{eqnarray*}
(r(\x))_k&=& (r(\e))_k \ , \ \ k\le \frac{1}{2}(N+1)(N+2),\\ 
(r(\s(\x)))_k&=& (r(\s(\e)))_k \ , \ \ k\le \frac{1}{2}N(N+1),\\
& \cdots  &\\
(r(\s^l(\x)))_k&=& (r(\s^l(\e)))_k \ , \ \ l\le N, \ k\le 
\frac{1}{2}(N+1-l)(N+2-l),\\
& \cdots  &
\end{eqnarray*}
Take $N=2M.$ Denote $N_l=\frac{1}{2}(N+1-l)(N+2-l)$. Then for $l\le M$, we have
\begin{eqnarray*}
d_1(r(\s^l(\x)), r(\s^l(\e)))&=&\sum^{+\infty}_{k=0}\frac{1}{2^k}
|(r(\s^l(\x)))_k-(r(\s^l(\e)))_k|\\ 
&  \le   &\sum^{+\infty}_{k=N_l+1}\frac{1}{2^k}=\frac{1}{2^{N_l}}\le 
2^{-\frac{1}{2}(M+1)(M+2)}<\frac{1}{2^M} \ ,
\end{eqnarray*}
therefore
$$
\r_1(R(\x), R(\e))\le 
\sum^{M}_{l=0}\frac{1}{2^l}\frac{2^{-M}}{1+2^{-M}}
+\frac{1}{2^M}<\frac{3}{2^M} \ ,
$$
which implies that $R: \S \to R(\S)$ is continuous. 

Take $\S^*=R(\S)$. Let $\tilde \x, \tilde \e \in\S^*$, then 
$\exists~ \x, \e \in \S,$ such that 
\begin{eqnarray*}
\tilde \x &=& (r(\x), r(\s(\x)), \cdots,  r(\s^k(\x)), \cdots),\\
\tilde \e &=& (r(\e), r(\s(\e)), \cdots,  r(\s^k(\e)), \cdots)
\end{eqnarray*}
Let $N=\frac{1}{2} (2k+1)(2k+2)$, when 
$$
\r_1 (\tilde \x, \tilde \e)< \frac{1}{2^{2N+1}},
$$
from Lemma~\ref{lem_metric} we have 
$$
d_1(r(\s^i(\x)), r(\s^i(\e))) < \frac{1}{2^N}, ~ i \le N.
$$
We get $\x_{ij}=\e_{ij}, ~i \le k,~j\le 2k$, and therefore 
$d_1(\x_i, \e_i) \le \frac{1}{2^{2k}},~i \le k$. 
Again from Lemma~\ref{lem_metric} we have 
$$
\r_1(R^{-1}(\tilde \x),R^{-1}(\tilde \e)) < 
\frac{1}{2^{2k-1}}+\frac{1}{2^k}.
$$
Therefore $R^{-1}: \S^* \to \S$ is also continuous,
hence $R$ is a homeomorphism. And $\forall \x \in \S,$
$$
R\s (\x)=(r(\s (\x )), r(\s ^2(\x )), \cdots, r(\s ^k(\x )), \cdots)=\s R(\x ),
$$
so $\s|_{\S} \sim \s|_{\S^*}$, while the subshift $(\S^*, \s)$ satisfies the 
condition (\ref{condition}).   \hfill  $\Box$

\begin{lem}  \label{lem_factor}
The full shift $(\S(I_1), \s)$ is a faithful representation of $(\S(I_0), \s)$. 
And $\forall ~(\S, \s) \le (\S(I_1), \s)$ with $\S$ compact, 
there exists a subshift  $(\S', \s) \le (\S(I_0), \s)$, such that 
$\s|_{\S} \sim \s|_{\S'}$.
\end{lem} 
\proof
$\forall a,b \in [0, 1]$,
$$
d_0(a, b) \le \frac{1}{2}d_1(a,  b).
$$
So $\forall \x, \e  \in \S(I_1)$,
$$
\r_0(\x, \e)\le \r_1(\x, \e),
$$
hence the identity mapping $i: \S(I_1)\rightarrow \S(I_0)$ is continuous. 
 Moreover, the following diagram 
commutes:
$$
\begin{array}{ccc}
 \S(I_1) & \mathop{\longrightarrow}\limits^{\s} & \S(I_1) \\
i\downarrow &  & \downarrow i \\
\S(I_0) & \mathop{\longrightarrow}\limits^{\s} & \S(I_0)  
\end{array}
$$
So $(\S(I_1), \s)$ is a faithful representation of  $(\S(I_0), \s)$.

$\forall ~ (\S, \s) \le (\S(I_1), \s)$ with $\S$ compact, 
$i: \S \to \S'=i(\S)=\S \subseteq \S(I_0)$ is a homeomorphism, therefore 
$(\S, \s)\le (\S(I_0),\s)$ and $\s|_{\S} \sim \s|_{\S'}$.   \hfill      $\Box$

The result in Lemma~\ref{lem_factor} can be generalized to:

\begin{pro}
If $(\S,\s)$ is a compact subshift of a faithful representation of $(X, f)$ under 
coding $h$, then $(\S, \s)$ is a conjugate representation of $(h(\S), f)$, a compact 
subsystem of $(X, f)$.   \hfill   $\Box$
\end{pro}

\begin{rem}  \label{rem_factor2}
$ \forall f \in M(I), \ \ f|_{I_0}$ is a factor of  $f|_{I_1}$.
\end{rem} 

For the identity mapping $i: I_1 \rightarrow I_0$, from the proof of Lemma 
\ref{lem_factor}, we have
$$
d_0(i(x), i(y))=d_0(x, y)\le \frac{1}{2}d_1(x, y), \ \ \forall x, y \in I_1 \ .
$$
So $i: I_1 \rightarrow I_0$ is continuous. 

$\forall f \in M(I)$, it is obvious that the following diagram 
commutes:
$$
\begin{array}{ccc}
 I_1 & \mathop{\longrightarrow}\limits^{f} & I_1 \\
i\downarrow &  & \downarrow i \\
I_0 & \mathop{\longrightarrow}\limits^{f} & I_0  
\end{array}
$$
So $f|_{I_0}$ is a factor of  $f|_{I_1}$.

\section{Conjugate Representations} \label{conjugate}

As discussed in \cite{Adler}, representing a general dynamical system by a 
symbolic one involves a fundamental complication: it is difficult and 
sometimes impossible to find 
a coding of a continuous one-to-one correspondence between orbits and 
symbolic sequences, and especially when we desire the shift system to be 
one of finite type and with a finite (or at most 
countable \cite{Fu92, Fu96, Kitchens}) alphabet. 
In this section, 
we will discuss conjugate representations of general continuous maps through general 
subshifts, which usually involve an uncountable alphabet.  

For a metric space $(X,d)$, there exists a huge number of 
continuous self-maps on 
$X$. The dynamics on $X$ is therefore diverse. 
The following theorem shows that 
the general subshifts of $(\S(X), \s)$ can be used as a unified
model for all continuous self-maps on the space $X$.

\begin{thm} \label{th_conj_rep}
Let $(X, d)$ be a metric space, 
$\forall f \in C(X), \exists (\S, \s) \le (\S(X), \s)$, 
such that $ (\S, \s)$ is a conjugate representation of $(X, f)$.
\end{thm} 

\proof  
Define a mapping $h: X \rightarrow \S(X)$ as
$$
h(x)=(x, f(x), f^2 (x), \cdots, f^n (x), \cdots),
$$
then $h$ is a one-to-one mapping.

Since $f$ is continuous,  for any 
positive integer $N$ and any sequence $\{ x_n\} \subset X$ with 
$ \lim_{n\rightarrow +\infty }x_n=x_0 \in X$, we have
$$
\lim_{n\rightarrow +\infty }d(f^k(x_n), f^k(x_0))=0, \ \  k=0, 1, \cdots, N.
$$
{}From
\begin{eqnarray*}
0 &  \le    & \r(h(x_n), h(x_0))=\sum^{+\infty}_{i=0} \frac{1}{2^i} 
\frac{d(f^i(x_n), f^i(x_0))}{1+d(f^i(x_n), f^i(x_0))}\\
& \le  & \sum^{N}_{i=0} \frac{1}{2^i} 
\frac{d(f^i(x_n), f^i(x_0))}{1+d(f^i(x_n), f^i(x_0))}+\sum^{+\infty}_{i=N+1} 
\frac{1}{2^i} \ ,
\end{eqnarray*}
we have 
$$
0 \le \lim_{n\rightarrow +\infty}\r(h(x_n), h(x_0)) \le \frac{1}{2^N} \ , 
 \ \forall N \ge 1,   
$$
that is, $h: X \rightarrow \S(X)$ is continuous.

Suppose we have a sequence 
$\{ \x^{(n)}=(\x^{(n)}_0,\x^{(n)}_1,\cdots,\x^{(n)}_i), \cdots \} \subset h(X)$ 
with
$$
\lim_{n\rightarrow +\infty}\x^{(n)}=\e=(\e_0,\e_1,\cdots, \e_i, \cdots) \in 
h(X),
$$
then
$$
\lim_{n\rightarrow +\infty}\r(\x^{(n)}, \e)=0.
$$
{}From
$$
0 \le \frac{d(\x^{(n)}_0, \e_0)}{1+d(\x^{(n)}_0, \e_0)} \le \r(\x^{(n)}, \e) 
\rightarrow 0 \mbox{ as } n \rightarrow +\infty,
$$
we obtain
$$
\lim_{n\rightarrow +\infty}d(\x^{(n)}_0, \e_0)=0.
$$
So $h^{-1}: h(X)\rightarrow X$ is also continuous, and therefore $h$ is a 
homeomorphism from $X$ to $h(X)$.

Moreover, 
$$
hf(x)=(f(x), f^2(x), \cdots)=\s h(x), \ \ \forall x \in X,
$$
i.e., $ hf=\s h. $

Take $\S =h(X)$, then $\S $ is invariant for the shift map $\s $, and $\S $ 
is a closed subset of $\S (X)$. So $(\S, \s) \le (\S(X), \s)$, and $f$ is 
topologically conjugate to $\s|_{\S}$.
\qed

\begin{rem}
The theorem above shows how to symbolize $(X, f)$ to $(\S, \s)$. 
Although the phase space of the latter may be more complex than that of the 
former, the map action of the latter is definitely simpler than that of the 
former. If we can understand the construction of the space $\S$ by some means, 
then the dynamics of $f$ on $X$ can be known accordingly. 
\end{rem}

We note that the subshift $(\S, \s)$ in the proof of Theorem \ref{th_conj_rep} 
satisfies the condition (\ref{condition}), i.e.,
$$
\forall \ x=(x_0, x_1,\cdots ), \ y=(y_0,y_1, \cdots )\in \S, \ x=y \mbox{ if 
and only if } x_0 = y_0 
$$
This is a restriction under which Theorem \ref{th_conj_rep} is reversible, as stated
below.

\begin{thm} \label{th_conj_rep2}
Suppose $(X, d)$ is a compact metric space, $(\S, \s) \le (\S(X), \s)$, and 
$(\S, \s)$ satisfies condition (\ref{condition}); then there exists a compact subset 
$X_0 \subseteq X$, and a continuous self-map $f$ on $X_0$, such that $(\S, \s)$
 is a conjugate representation of $(X_0, f)$.
\end{thm} 
\proof
Let $(\S, \s)$ be a subshift of $(\S(X), \s)$ satisfying condition (\ref{condition}). 
Define $\f: \S \rightarrow X$ as
$$
\f((x_0, x_1,\cdots))=x_0, \ \ \forall \ (x_0, x_1,\cdots) \in \S.
$$
Then $\f$ is a one-to-one mapping. Denote $\f(\S)$ by $X_0$. $\forall (x_0, 
x_1,\cdots), (y_0,y_1, \cdots)  \in  \S,$
\begin{eqnarray*}
d(\f((x_0, x_1,\cdots)), \f((y_0,y_1, \cdots))) & = & d(x_0, y_0)\\
& < & (1+d(x_0, y_0))\r((x_0, x_1,\cdots), (y_0,y_1, \cdots)),
\end{eqnarray*}
$\forall \ep > 0, $  let $\d=\ep /(1+\ep)$, then $0 < \d < 1.$ Whenever 
$\r((x_0, x_1,\cdots), (y_0,y_1, \cdots)) < \d$,
\begin{eqnarray*}
\frac{d(x_0,y_0)}{1+d(x_0,y_0)} & < & \r((x_0, x_1,\cdots), (y_0,y_1, \cdots)) < 
\d, \\
d(x_0, y_0) & < & \frac{\d}{1-\d},  \\
d(\f((x_0, x_1,\cdots)), \f((y_0,y_1, \cdots))) & < & (1+\frac{\d}{1-\d})\d=\ep,
\end{eqnarray*}
so $\f: \S \rightarrow X_0$ is continuous.

Since $X$ is compact, $\S(X)$ is also compact and so is $\S$. As a
result, $\f: \S \rightarrow X_0$ is a homeomorphism and hence $X_0 \subseteq X$ 
is compact. 
Take $f: X_0 \rightarrow X_0 $ as $f=\f \s \f^{-1}$, then $f$ is continuous, and 
$f|_{X_0} \sim \s|_{\S}$.
\qed

Naturally, one would like to ask if Theorem \ref{th_conj_rep2} still holds for 
subshifts  $(\S, \s)$ which don't satisfy the condition (\ref{condition})? The answer 
is yes if one can construct another subshift $(\S^*, \s)$ such that $(\S, \s) 
\sim (\S^*, \s)$, and $(\S^*, \s)$ satisfies the condition (\ref{condition}). This can 
be done at least for one-dimensional self-maps, as shown in the following 
example.

\vspace{0.3cm}
\noindent {\sc Example}  \  Let $X=[0, 1], d(x,y)=|x-y|, \S=\S(2)=\{(x_0, x_1, 
\cdots, x_i, \cdots): x_i=0, 1, i \ge 0 \}$. Take $\S^*=R(\S)=\{R(x): x \in 
\S \}$, where $R: \S \rightarrow \S(X)$ is defined as 
\begin{eqnarray*}
R(x)     & = & (r_{10}(x), r_{10}(\s(x)), \cdots, r_{10}(\s^k(x)), \cdots),\\
r_{10}(x)& = & 0.x_0x_1 \cdots x_k \cdots  \ \ \mbox{(decimal fraction)}, 
\forall x=(x_0, x_1, \cdots, x_k, \cdots) \in \S,
\end{eqnarray*}
then it can be verified (see Remark~\ref{rem_embed} in 
Section~\ref{quasi} for detail) that $R: \S 
\rightarrow \S^*$ is a topological conjugacy. So $\s|_{\S} \sim \s|_{\S^*}$, 
while $(\S^*, \s)$ satisfies the condition (\ref{condition}).
\vspace{0.3cm}

In the following theorem, we reach a more general 
conclusion for the case of $X=[0,1]$. That is, when $X=[0,1]$, Theorem 
\ref{th_conj_rep2} 
holds for all subshifts $(\S, \s)\le (\S(I_0), \s)$ with $\S$ compact 
in $(\S(I_1), \s)$, regardless of the condition (\ref{condition}).

\begin{thm} \label{th_symb_rep}
$\forall ~ (\S, \s) \le (\S(I_0), \s)$ with $\S$ compact in $\S(I_1)$, 
there exists an $f \in C(I_0)$ 
and $\L \subseteq I_0$ with $\L$ compact and invariant for $f$, 
such that $(\S, \s)$ is a conjugate representation  of $(\L, f)$.
\end{thm} 
\proof
Since $\S$ is compact in $\S(I_1)$, $(\S, \s)$ is also a subshift of 
$(\S(I_1), \s)$.
{}From Lemma \ref{lem_condition}, there exists a subshift $(\S^*, \s)$ of 
$(\S(I_1), \s)$ with $\S^*$ also compact, such that 
$\s|_{\S} \sim \s|_{\S^*}$, and $(\S^*, \s)$ satisfies the 
condition~(\ref{condition}).

{}From Lemma~\ref{lem_factor}, $(\S^*, \s)$ is also a subshift of 
$(\S(I_0), \s)$ satisfying the condition~(\ref{condition})

{}Similar to the proof of Theorem \ref{th_conj_rep2}, it can be shown that 
there exists a compact subset $\L \subseteq I_0$, and a 
continuous self-map $f$ on $\L$ such that $(\S^*, \s)$ is a conjugate 
representation of $(\L, f)$, a subsystem of $(I_0, f)$. Therefore $(\S, \s)$ is 
a conjugate representation of $(\L, f)$.
\qed

Putting Theorem \ref{th_conj_rep} and Theorem \ref{th_symb_rep} together, we have the 
following representation theorem for one-dimensional continuous self-maps.

\begin{thm} \label{th_rep}
 $\forall f \in C(I_0), \exists (\S, \s)\le 
(\S(I_0), 
\s)$, such that $(\S, \s)$ is a conjugate representation of $(I_0, f)$; 
conversely, $\forall (\S, \s) \le (\S(I_0), \s)$ with $\S$ compact in 
$\S(I_1), \exists f \in C(I_0)$ and $\L \subseteq I_0$, where $\L$ is 
compact and invariant for $f$, such that 
$(\S, \s)$ is a conjugate representation  of $(\L, f)$.   \hfill   $\Box$
\end{thm}

\section{The Second Order Representations}\label{sec_second}

Theorem \ref{th_conj_rep} indicates that $\forall f \in C(X), 
(X, f)$ can be embedded 
in $(\S(X), \s)$, denoted by $(X, f)\hookrightarrow (\S(X), \s)$. This 
embedding relation means that the system $(X, f)$ may be represented by a 
subshift of  $(\S(X), \s)$. Naturally, one may further consider the 
representation of the system  $(\S(X), \s)$ itself. To do so, 
one can regard $\S(X)$ as a symbol space and denote by $\S^2(X)$ the symbol 
sequence space 
$\S(\S(X))$, i.e., 
$$
\S^2(X)=\{x=(x_0, x_1, \cdots, x_i, \cdots): x_i\in \S(X), i\ge 0 \}.
$$
We specify a metric $\r^{(2)}$ on $\S^2(X)$ by
$$
 \r^{(2)}(x, y)=\sum^{+\infty}_{k=0} \frac{1}{2^k} 
\frac{\r(x_k, y_k)}{1+\r(x_k, y_k)},
$$
where $\r$ is the metric on $\S(X)$. We denote by $\s_2$ the shift map on 
$\S^2(X)$, and if no confusion caused, we also denote by $\s$ the shift map 
on $\S^2(X)$. And so we get a general symbolic dynamics system 
$(\S^2(X), \s)$. Similarly, we can define general symbolic dynamics system 
$(\S^k(X), \s_k)$ (or denote by $(\S^k(X), \s)$ if no confusion caused) for 
$k\ge 3,$ and call it the $k$-th order symbolic representation of 
dynamics on $X$. 

For higher order symbolic sequence spaces, we have the following 
general result.

\begin{thm} \label{th_homeo}
Suppose $(X, d)$ is a metric space. Then $\S^2(X)$ is homeomorphic to $\S(X)$.
In general, for $k\ge 2,~ \S^k(X)$ is homeomorphic to $\S^{k-1}(X)$.
\end{thm}
\proof
For $\tilde x, \tilde y \in \S^2(X)$, denote
\begin{eqnarray*}
\tilde x = (\tilde x_0, \tilde x_1, \cdots,\tilde x_i, \cdots), \ \ 
\tilde y = (\tilde y_0, \tilde y_1, \cdots,\tilde y_i, \cdots);
\end{eqnarray*}
and 
\begin{eqnarray*}
\tilde x_i = (x_{i0}, x_{i1}, \cdots, x_{ij}, \cdots), \ \
\tilde y_i = (y_{i0}, y_{i1}, \cdots, y_{ij}, \cdots),
\end{eqnarray*}
where $x_{ij}, y_{ij} \in X$.

Define $h: \S^2(X) \to \S(X)$ as:
$$
h(\tilde x) =
(x_{00}x_{10}x_{01} \cdots x_{i0}x_{i-1,1}\cdots x_{1,i-1}x_{0i}\cdots),
$$
then $h$ is one-to-one and onto.

When 
$$
\r^{(2)}(\tilde x , \tilde y) < \frac{1}{2^{2(2N+1)+1}},
$$
from Lemma~\ref{lem_metric}, we have 
$$
\r(\tilde x_i, \tilde y_i) < \frac{1}{2^{2N+1}}, ~ i \le 2N+1,
$$
therefore
$d(x_{ij}, y_{ij}) < 2^{-N}, ~ i\le 2N+1,~ j\le N$, so we get 
$$
\r(h(\tilde x), h(\tilde y)) < \frac{1}{2^{N-1}} + \frac{1}{2^{N_1}},
$$
where $N_1= \frac{1}{2} (N+1)(N+2) + N+1$.

On the other hand, let $x=(x_0x_1 \cdots x_k \cdots), 
y=(y_0y_1 \cdots y_k \cdots) \in \S(X)$, 
when $\r(x,y)< 2^{-(2N+1)}$, where we suppose 
$N=\frac{1}{2} (2k+1)(2k+2)$, we have $d(x_i, y_i)< 2^{-N}, ~ i \le N$.
Let
\begin{eqnarray*}
h^{-1}(x)=\tilde x = (\tilde x_0, \tilde x_1, \cdots,\tilde x_i, \cdots),
      &  & \tilde x_i=(x_{i0}, x_{i1}, \cdots, x_{ij}, \cdots),\\
h^{-1}(y)=\tilde y = (\tilde y_0, \tilde y_1, \cdots,\tilde y_i, \cdots),
&  & \tilde y_i = (y_{i0}, y_{i1}, \cdots, y_{ij}, \cdots),
\end{eqnarray*}
then $d(x_{ij}, y_{ij}) < 2^{-N}, ~ i+j \le 2k+1$, therefore 
$d(x_{ij}, y_{ij}) < 2^{-N}, ~ i\le k,~ j \le k$. This implies 
$\r(\tilde x_i, \tilde y_i) < 2^{1-N} + 2^{-k},~i \le k$, therefore 
$\r^{(2)}(\tilde x, \tilde y)< 2^{2-N}+2^{1-k} + 2^{-k}$, i.e., we have 
$$
\r^{(2)}(h^{-1}(x), h^{-1}(y)) < 
\frac{1}{2^{N-2}} + \frac{1}{2^{k-1}} + \frac{1}{2^{k}}.
$$
Hence both $h$ and $h^{-1}$ are continuous, and therefore 
$\S^2(X)$ is homeomorphic to $\S(X)$. 

Similarly, we can check that for $k\ge 2,~ \S^k(X)$ is homeomorphic to 
$\S^{k-1}(X)$.
\qed

On the other hand, define $\f: \S(X)\rightarrow \S^2(X)$ as 
$$
\f(x)=(x, \s(x), \cdots, \s^k(x), \cdots),
$$
then if $X$ is compact,  $\f$ can be verified to be a topological conjugacy 
from $\S(X)$ to $\f(\S(X))\subseteq \S^2(X)$. So $(\S(X), \s)\hookrightarrow 
(\S^2(X), \s)$.

In general, for a compact metric space $X$, we have the following embedding 
sequence:
\begin{eqnarray}
(X, f)\hookrightarrow (\S(X), \s)\hookrightarrow (\S^2(X), \s)\hookrightarrow 
\cdots \hookrightarrow (\S^k(X), \s)\hookrightarrow \cdots        \label{3.1}
\end{eqnarray}

That's why we discuss the higher order representations. And the topic is also 
motivated by Nasu (\cite{Nasu}) and is helpful to study maps in symbolic 
dynamics discussed in \cite{Nasu}, as well.

%
%

As we discussed earlier that $(X, f)\hookrightarrow (\S(X), \s)$ 
shows a transformation between dynamics on $X$ and subshifts structure in 
$(\S(X), \s), (\S^k(X), \s)\hookrightarrow (\S^{k+1}(X), \s)$ indicates that the 
subshifts structure in $(\S^k(X), \s)$ can be transformed to the subshifts 
structure in $(\S^{k+1}(X), \s)$. In other words, the symbolic 
dynamics system $(\S^k(X), \s)$, and its subshifts, can be further represented 
by the one order higher symbolic dynamics system $(\S^{k+1}(X), \s)$.  
If the embedding sequence (\ref{3.1}) is finite, then the above transformation 
or representations will not go on without limit. This means that the piling up 
of symbol sequence spaces $\S^k(X)$ will not cause an unlimited increase of the 
complexity of the corresponding shift systems. In particular, if we have 
$(\S(X), \s)\sim (\S^2(X), \s)$, then $(\S(X), \s)$ is an ultimate 
(or final) representation. We may ask  under what conditions does 
$(\S(X), \s)\sim (\S^2(X), \s)$ hold?  We have the following theorems.

\begin{thm} \label{th_iff}
$(\S^2(X), \s)$ is topologically conjugate to $(\S(X), \s)$ 
if and only if $\S(X)$ is homeomorphic to $X$. 
In general, for $k\ge 1,~(\S^{k+1}(X), \s)\sim (\S^k(X), \s)$ 
if and only if $\S^k(X)$ is  homeomorphic to $\S^{k-1}(X)$.
\end{thm}

\proof
Suppose $\a: \S(X)\to X$ is a homeomorphism. Define $\f: \S^2(X)\to \S(X)$ as:
$$
\f(\x)=(\a(\x_0),\a(\x_1),\cdots,\a(\x_k),\cdots), ~ 
\x=(\x_0,\x_1,\cdots,\x_k,\cdots) \in \S^2(X),
$$
then $\f$ is a homeomorphism, and $\f \s_2 =\s \f$. Therefore 
$(\S^2(X), \s)\sim (\S(X), \s)$.

Conversely, let $\f: \S^2(X)\to \S(X)$ is a topological conjugacy.
$\forall ~ x \in \S(X)$, denote $\tilde x = (x,x,\cdots,x,\cdots)$, 
then $\tilde x$ is a fixed point of $\s_2$ in $\S^2(X)$. 
Since $\s_2 \sim \s$, $\f(\tilde x)$ is also a fixed point of $\s$ 
in $\S(X)$. So $\exists ~ x^* \in X$, such that 
$\f(\tilde x)=(x^*,\cdots, x^*, \cdots)$. Define $\a: \S(X)\to X$ as: 
$\a(x)=x^*$. Then $\a$ is a homeomorphism.

Similarly, we can prove the general case of the Theorem.
\qed

{}From Theorems~\ref{th_homeo}~and~\ref{th_iff}, 
the following corollary is immediate:

\begin{cor} \label{cor_3-rep}
Suppose $(X, d)$ is a metric space, 
then we always have 
$$
(\S^k(X), \s)\sim (\S^2(X), \s),~\forall ~ k \ge 3.     
$$ \hfill $\Box$            
\end{cor}

So the embedding sequence~(\ref{3.1}) is finite,   
and the third or higher order representations are therefore not necessary. 

The following results show that we further have 
$(\S(X), \s)\sim (\S^2(X), \s)$ when $X=I_1$.

\begin{thm} \label{th_high-order2}
$(\S(I_1), \s)$ is topologically conjugate to $(\S^2(I_1), \s)$, 
i.e., $(\S(I_1), \s)\sim (\S^2(I_1), \s)$. 
\end{thm} 

\proof
$\forall (x_0, x_1, \cdots, x_k, \cdots)\in \S^2([0, 1])$, where
$$
x_i\in \S([0, 1]), \ x_i=(a_0^{(i)},a_1^{(i)}, \cdots, a_k^{(i)}, \cdots), \ 
a_k^{(i)} \in [0, 1], \ i\ge 0,\ k\ge 0.
$$
Rewrite $a_k^{(i)}$ by
$$
a_k^{(i)}=r_2(a_k^{(i)})=\sum^{+\infty}_{j=0}2^{-(j+1)}a_{kj}^{(i)} \ , \ \ 
i,k\ge 
0,
$$
then define $\a: \S([0, 1])\rightarrow [0, 1]$ as:
$$
\a(x_i)=\sum^{+\infty}_{l=0}\displaystyle\sum^l_{{j=0 \atop k+j=l}\atop k\ge 
0}2^{-(\frac{1}{2}l(l+1)+j+1)}a^{(i)}_{kj} \ , \ \ i\ge 0,
$$  
So $\a(x_i)$ is also a binary fraction. Further,  $\a: \S([0, 1])\rightarrow [0, 
1]$ 
is a one-to-one and onto mapping. Choose $d_1$ as the metric on $[0, 1]$, and 
correspondingly $\r_1$ as the metric on $\S([0, 1])$. Then $\forall x_i, y_i \in 
\S(I_1)$, 
\begin{eqnarray*}
x_i = (a^{(i)}_0, a^{(i)}_1, \cdots, a^{(i)}_k, \cdots),  \ \ 
y_i = (b^{(i)}_0, b^{(i)}_1, \cdots, b^{(i)}_k, \cdots),
\end{eqnarray*}
rewrite $a^{(i)}_k$ and $b^{(i)}_k$ as
\begin{eqnarray*}
a^{(i)}_k &=& r_2(a_k^{(i)}) = \sum^{+\infty}_{j=0}2^{-(j+1)}a_{kj}^{(i)} \ ,\\
b^{(i)}_k &=& r_2(b_k^{(i)}) = \sum^{+\infty}_{j=0}2^{-(j+1)}b_{kj}^{(i)} \ .
\end{eqnarray*}
Whenever
$$
\r_1(x_i, y_i)< \frac{1}{2^{2N+1}},
$$
we have
$$
d_1(a^{(i)}_k, b^{(i)}_k)< \frac{1}{2^N}, \ \ k\le N,
$$
therefore
$$
a_{kj}^{(i)}=b_{kj}^{(i)}, \ \ k\le N, \ j\le N-1.
$$
So we have
$$
d_1(\a(x_i), \a(y_i))\le 2^{1-\frac{1}{2}N(N+1)},
$$
namely, $\a: \S(I_1)\rightarrow I_1$ is continuous.

Similarly, we can check that $\a^{-1}: I_1 \to \S(I_1)$ is also continuous. 
Hence $\S([0, 1])$ is homeomorphic to $[0, 1]$. 
From Theorem~\ref{th_iff} we have 
$$ 
(\S(I_1), \s)\sim (\S^2(I_1), \s).
$$
The Theorem is proved.
\qed

We may ask if we also have $(\S(I_0), \s)\sim (\S^2(I_0), \s)$? We guess this 
is not true but only have:

\begin{pro}
$(\S(I_1), \s)$ is a faithful representation for both 
$(\S(I_0), \s)$  and  \\ $(\S^2(I_0), \s)$.
\end{pro}

\proof
{}From Lemma \ref{lem_factor},  $(\S(I_1), \s)$ is a faithful representation of 
$(\S(I_0), \s)$.

Since $d_0(\a, \b)\le \frac{1}{2}d_1(\a, \b), \ \ \forall \a, \b \in [0, 1]$, we 
have
\begin{eqnarray*}
\r_0(a, b)& \le & \r_1(a, b), \ \ \forall a, b \in \S([0, 1]),\\
\r^{(2)}_0(x, y)& \le & \r^{(2)}_1(x, y), \ \ \forall x, y \in \S^2([0, 1]).
\end{eqnarray*}
Similar to the proof of Lemma \ref{lem_factor}, 
we can prove that $(\S^2(I_1), \s)$ 
is topologically semi-conjugate to $(\S^2(I_0), \s)$ under the identity mapping 
$i: \S^2(I_1)\rightarrow \S^2(I_0)$. Therefore 
$(\S^2(I_1), \s)$ is a faithful representation of $(\S^2(I_0), \s)$.

So $(\S(I_1), \s)$ is a faithful representation for both $(\S(I_0), \s)$ 
and $(\S^2(I_0), \s)$.          
\qed

Note that from Theorem~\ref{th_iff}, 
$$
(\S^2(N), \s) \nsim (\S(N), \s).
$$ 
We suspect that 
$$(\S^2(I_0), \s) \nsim (\S(I_0), \s).
$$ 
And we also suspect in most circumstances 
$$
(\S^2(X), \s) \nsim (\S(X), \s).
$$
So it is very necessary to study the second order representations.

$(\S^2(N), \s) \nsim (\S(N), \s)$ also means that shift maps with a finite or 
countable alphabet is not sufficient for the study of symbolic 
representations of dynamical systems, and so it is necessary to study 
symbolic dynamics with an uncountable alphabet.

\section{Quasi-Representations}  \label{quasi}

While we have shown the existence of the topological conjugacy between a 
one-dimensional map and a subshift of a symbolic dynamics system, in practice 
it is often difficult to construct such conjugacy for applications. Instead, it 
may be easier to find a topological semi-conjugacy, which sometimes is 
sufficient for the problems under investigation. In this section we discuss 
quasi-representations of one-dimensional dynamical systems. 

The symbolic dynamics system $(\S(I), \s)$ is an extension of the usual 
symbolic dynamics system $(\S(N), \s)$. Equip $\S(N)$ with a metric $\r$ as 
follows:
$$
\r(x, y)=\sum^{+\infty}_{i=0} \frac{1}{2^i} 
\frac{|x_i-y_i|}{1+|x_i-y_i|}.
$$
Take from $I_1$ a convergent sequence $\{a_k\}$ with the limit $a \in I_1$, 
where $a_i \ne a_j,~ \forall~ i\ne j$. Let
\begin{eqnarray*} 
S_{\infty} & = & \{a_1,a_2, \cdots, a\}, \\
\S_{\infty} & = & \{(x_0, x_1, \cdots) \in \S(I_1): x_i \in S_{\infty}, i \ge 
0\},
\end{eqnarray*}
then $\S_{\infty}$ is compact in $\S(I_1)$, and 
$(\S_{\infty}, \s)\le (\S(I_0), \s)$. From Theorem \ref{th_symb_rep}, 
$\exists f \in C(I_0)$ and $\L \subseteq I_0, \L $ is compact and invariant 
for $f$, such that $f|_{\L} \sim \s|_{\S_{\infty}}$.

$\forall (\S, \s) \le (\S(N), \s)$, we have
$$
{h_{\rm top}}(\s|_{\S}) \le {h_{\rm top}}(\s|_{\S(N)})=\log N 
 \ne{h_{\rm top}}(\s|_{\S_{\infty}})=+\infty,
$$ 
So $\s|_{\S} \not\sim 
\s|_{\S_{\infty}}$, and $f|_{\L} \not\sim \s|_{\S}$. Therefore the following 
result is obvious:

\begin{thm} \label{th_not_conj}
There exists an $f \in C(I_0)$, such that 
$\forall (\S, \s) \le (\S(N), \s), f|_{\L} 
\not\sim \s|_{\S}$, where $\L \subseteq I_0$ is a closed invariant subset for 
$f$.          \hfill     $\Box$
\end{thm}  

Theorem \ref{th_not_conj} indicates once again that the scope of application 
for $(\S(N), \s)$ is 
quite limited. So it is necessary to study its extensions, such as $(\S(I_0), 
\s)$, etc. Nevertheless, the system $(\S(N), \s)$ still has its special 
significance. For example, $(\S(N), \s)$ can be used effectively to characterize 
the dynamics of multimodal one-dimensional maps (see \cite{Hao89} and the 
references therein). The significance of $(\S(N), \s)$ can also be shown by the 
following theorem.

\begin{thm} \label{th_quasi}
 $\forall f \in C(I_0), \exists (\S, \s) \le 
(\S(N), \s),   N \ge 2,$ such that $(\S, \s)$ is a quasi-representation of 
$(I_0, f)$.
\end{thm} 
\proof
First, we prove $(\S(I_0), \s)$ is topologically semi-conjugate to $(\S(N), 
\s)$.

Let
$$
[0, 1]=\bigcup^{N-1}_{k=0}A_k, \ \ \mbox{where}\   A_0=[0, \frac{1}{N}], 
A_k=(\frac{k}{N}, \frac{k+1}{N}], \ k=1, 2, \cdots, N-1. 
$$
Define $\a: [0, 1]\rightarrow \{0, 1, \cdots, N-1\}$ as $\a(a)=k, \ a\in A_k.$ 
And define $\f: \S(I_0) \to \S(N)$ as:
$$
\f((x_0, x_1, \cdots, x_k, \cdots))=(\a(x_0), \a(x_1), \cdots, \a(x_k), \cdots).
$$
Then
$$
|\a(x_i)-\a(y_i)|=|k-l|, \ \ x_i \in A_k, \ y_i\in A_l.
$$
If $k\ne l$, we have
$$
\frac{1}{N}\le |x_i-y_i|\le 1, \ \ 1\le |\a(x_i)-\a(y_i)|\le N-1,
$$
therefore
$$ 
\frac{|\a(x_i)-\a(y_i)|}{1+|\a(x_i)-\a(y_i)|}  < N^2 
\frac{|x_i-y_i|}{1+|x_i-y_i|}.
$$
If $k=l$, it is obvious that
$$
\frac{|\a(x_i)-\a(y_i)|}{1+|\a(x_i)-\a(y_i)|}\le N^2 
\frac{|x_i-y_i|}{1+|x_i-y_i|}.
$$
So $\r(\f(x), \f(y))\le N^2\r_0(x, y)$, thus $\f$ is continuous. $\f$ is also 
surjective. And the following diagram commutes:
$$
\begin{array}{ccc}
 \S(I_0) & \mathop{\longrightarrow}\limits^{\s} & \S(I_0) \\
\f \downarrow &  & \downarrow \f \\
\S(N) & \mathop{\longrightarrow}\limits^{\s} & \S(N)  
\end{array}
$$
So we have proved that $(\S(I_0), \s)$ is topologically semi-conjugate to 
$(\S(N), \s)$.

$\forall f\in C(I_0)$, there exists $(\S^*, \s)\le (\S(I_0), \s)$ such that 
$f|_{I_0}\sim \s|_{\S^*}$. Denote the topological conjugacy by $\p$. Then we 
can define $\l: I_0 \rightarrow \f(\S^*)$ as $\l=\f\p$,
$$
\begin{array}{rcccc}
 I_0 &\mathop{\longrightarrow}\limits^{\p}& \S^* & 
\mathop{\longrightarrow}\limits^{\f} 
& \f(\S^*) \\
f\downarrow &  & \downarrow \s& & \downarrow \s\\
I_0 & \mathop{\longrightarrow}\limits^{\p} & \S^*& 
\mathop{\longrightarrow}\limits^{\f}& 
\f(\S^*) 
\end{array}
$$
So $\l$ is a continuous and onto map, and $\l f=\s \l.$

Since $\s(\S^*)\subseteq \S^*, \s(\f(\S^*))\subseteq \f(\S^*)$. So $\f(\S^*)$ 
is invariant for $\s.$

$\S^*$ is closed, $\S(I_0)$ is compact, so $\f(\S^*)$ is closed. 
So $(\f(\S^*), \s) \le (\S(N), \s)$.

Take $\S=\f(\S^*)$, then $f|_{I_0}$ is topologically semi-conjugate to 
$\s|_{\S}.$
\qed

{}From the above discussions, we have the following remarks.
              
\vspace{0.3cm}

\begin{rem}
 $\exists (\S^*, \s)\le (\S(I_0), \s)$, such that 
$\forall (\S, \s)\le (\S(N), \s), \s|_{\S}\not\sim \s|_{\S^*}$.
\end{rem}

\begin{rem}
$(\S(N), \s)$ is a factor of $(\S(M), \s)$ when $M>N$. 
And $(\S(N), \s)$  is a factor of $(\S(I_0), \s)$.
\end{rem}

\begin{rem} \label{rem_embed}
$(\S(N), \s)$ can be embedded in $(\S(I_0), \s)$, 
i.e., $\exists (\S, \s)\le (\S(I_0), \s)$, such that $\s|_{\S(N)}\sim \s|_{\S}$  
  \ , where $N \ge 2.$
\end{rem}

In fact, the subshift $\S$ may be chosen as: $\S=R_N(\S(N))$, where 
\begin{eqnarray*}
R_N(x)     & = & (r_N(x), r_N(\s(x)), \cdots, r_N(\s^k(x)), \cdots),\\
r_N(x)& = &  \sum^{+\infty}_{i=0}  \frac{x_i}{N^{i+1}}, \ \forall x=(x_0, x_1, 
\cdots)\in \S(N).
\end{eqnarray*}
Then if $R_N(x)=R_N(y)$, we have $\forall k \ge 0,$
\begin{eqnarray*}
r_N(\s^k(x))&=&r_N(\s^k(y)),\\
r_N(\s^{k+1}(x))&=&r_N(\s^{k+1}(y)),
\end{eqnarray*}
These imply $x_k=y_k, \forall k \ge 0.$ So $x=y$, and $R_N: \S(N)\rightarrow \S$ 
is a one-to-one mapping. 

When $x, y \in \S(N)$ and $\r(x, y)<1/2^{M+1}$, we have $x_i=y_i, \ i=0, 1, 
\cdots, M.$ So
$$
|r_N(\s^k(x))-r_N(\s^k(y))|\le \sum^{+\infty}_{i=M-k+1}  
\frac{|x_{k+i}-y_{k+i}|}{N^{i+1}}<\frac{1}{N^{M-k}}, \ 0\le k\le M.
$$
Therefore
$$
\r_0(R_N(x), R_N(y))\le 
\sum^{M}_{k=0}\frac{1}{2^k}\frac{1}{N^{M-k}+1}+\sum^{+\infty}_{k=M+1} 
\frac{1}{2^k}.
$$
Since $N\ge 2, \ \ 1/N\le 1/2$,
$$
\r_0(R_N(x), R_N(y))<\frac{M+2}{2^M}.
$$
So $R_N$ is continuous, and therefore is a homeomorphism.

It is obvious that $R_N\s|_{\S(N)}=\s|_{\S}R_N.$ So $\s|_{\S(N)}\sim 
\s|_{\S}$. So we have the result in Remark~\ref{rem_embed}

\section{Partitions and Representations}  \label{partition}

In Section~\ref{quasi} quasi-representations of one-dimensional dynamical 
systems are discussed. From the proof of Theorem~\ref{th_quasi}, when the 
whole interval (phase space) is divided into some smaller pieces, and code 
each piece with a symbol, then we get a quasi-representation of the original 
system. This shows a direct connection between partitions and representations.
Partitions are natural ways to associate a symbolic
sequence with an orbit by tracking its history. 
As pointed out in \cite{Adler}, in order to get a useful symbolism, one needs 
to construct a partition with special properties. For example, when a partition 
is Markov, the sysytem can be represented by a subshift of finite type. Some 
hyperbolic systems, such as Anosov systems, axiom A systems, psuedo-Anosov 
systems, and hyperbolic automorphisms on $n$-tori, $n \ge 2$, admit Markov 
partitions. However, for non-hyperbolic systems, there may be no Markov 
partitions. Therefore other than Markov partitions, some more general 
partitions for these systems need to be used, so that if a certain kind of 
partition exists for a non-hyperbolic system, the system then can be 
represented by a subshift of infinite type.

In this section we generalize some concepts and main results discussed 
in \cite{Adler}.

The discussion below are for dynamical systems $(X, \f)$, where $X$ is a 
compact metric space with metric $ d ( \cdot \, , \cdot ) $ and
$\f $ is a homeomorphism of $X$ onto itself.

\begin{defn} \label{defn_top_part}
A finite or countable family of subsets 
$\MR = \{ R_{i},  i \in \MI \}$
is called a {\em topological partition} for a dynamical system $(X, \f)$ if:
{\rm (1)} each $R_{i}$ is open;
{\rm (2)} $R_{i} \cap R_{j} = \varnothing , i \ne j $;
{\rm (3)} $X = \bigcup_{i \in \MI}\overline{R_{i}}$;
{\rm (4)} $\forall i \in \MI, 
\card \{k \in \MI, \f R_i \cap R_k \ne \varnothing\}< + \infty$. 
\end{defn}

\begin{rem}
Other authors have taken the sets $R_i$ to be closed sets
with the property that each is the closure of its interior.  The 
variation introduced by Adler here is slightly more general, just 
enough to make some notation and certain arguments simpler. In fact, 
there is a example of 
partition whose elements are not the interiors of their closures 
{\rm (\cite{Adler})}.
\end{rem}

\begin{rem}
When $\card~\MI <+\infty$, then Definition~\ref{defn_top_part} coincides with 
the definition of topological partitions in {\rm \cite{Adler}}.
\end{rem}

Given two topological partitions
$$ \MR = \{ R_{i},  i \in \MI_1 \} ~ \mbox{and}~
 \MS = \{ S_{j},  j \in \MI_2 \},
$$
we  define their {\em common topological refinement}
$\MR \vee \MS $ as
$$
\MR \vee \MS = \{ {R_{i} \cap 
S_{j}} , i \in \MI_1, \ j \in \MI_2 \} .
$$

\begin{lem}
For dynamical system $(X , \f )$ with topological
partition $\MR$, the set $\f ^{n} \MR $ defined by
$$
\f ^{n} \MR = \{ \f ^{n} R_{i}, i \in \MI \}
$$
is also a topological partition; and $\forall m \le n, 
\bigvee _{k=m}^{n} \f ^{k} \MR$ is again a topological partition. 
\end{lem}

A topological partition is called a {\em generator} for a 
dynamical system $(X , \f)$ if 
$$\lim _{n \to \infty }
D \left ( \bigvee _{-n}^{n} \f ^{k} \MR \right ) = 0 .
$$
Where $D (\MR)$ denotes the {\em diameter} of a partition 
$\MR$, 
$$D(\MR) = \max _{R_{i} \in \MR}
D( R_{i}) $$
where $D (R_{i}) \equiv \sup _{x,y \in R_{i}} d (x , y).$

We say that a topological partition
$\MR$ for a dynamical system $(X , \f)$
satisfies
{\em the $n$-fold intersection property}
for a positive integer $n \ge 3$ if
$$
R_{s_{k}} \cap \f ^{-1} R_{s_{k+1}} \ne \varnothing , \
1 \le k \le n-1 \Rightarrow \bigcap _{k=1}^{n} \f ^{-k} R_{s_{k}} \ne 
\varnothing .
$$
Furthermore, we call a topological partition {\em Markov} if it
satisfies the $n$-fold intersection property for all $n \ge 3 .$

A homeomorphism $\f$ is said to
be {\em expansive} if there exists a real number $c>0$ such that
if $ d (\f ^{n} (x) , \f^{n} (y) ) < c$ for all $n \in \Z$,
then $x = y$.

Suppose a dynamical system $(X, \f)$ has a Markov generator
$\MR\!=\!\{ R_{i}, i \in \MI \} $.
We define an associated subshift of finite type  $(\S_{\MR}, \s)$ 
over a finite or countable alphabet by
$$\S_{\MR} = \{ s = (s_{n})_{n \in \Z} :
R_{s_{n-1}} \cap \f ^{-1} R_{s_{n}} \ne \varnothing ,
s_{n}\in \MI, n \in \Z \} . 
$$

Similar to Theorems 6.5 and 6.13 in \cite{Adler}, we have the following result.

\begin{thm}
If the dynamical system $(X , \f)$ is expansive and
has a Markov generator $\MR = \{ R_{i} , i \in \MI \}$,
then the map $\pi : \S_{\MR} \to X$ defined by
$$
\pi (s) =  \bigcap _{ n = 0}^{\infty }
\overline{\f^{n} R_{s_{-n}} \cap \f ^{n-1} R_{s_{-n+1}} \cap \dots 
\cap \f^{-n} R_{s_{n}}}  
$$
gives a regular representation $(\S_{\MR}, \s)$ of $(X, \f)$.
Moreover, a subshift of finite type is a semi-conjugate representation of an 
expansive dynamical system $(X, \f )$ if and only if $(X, \f)$ has 
a Markov partition.   \qed
\end{thm}

\section{Partial Representations}  \label{distillation}

It would be better, under certain conditions, to symbolize a dynamical 
system by a conjugate representation and using a finite or countable alphabet. 
This section will show that 
if there exists a distillation, then we can achieve this target, although 
we have to abandon the quest of representing all points in the phase space, 
instead, only represent an invariant subset, that is, we obtain a 
partial representation.  

In contrast with partitions, if there exist pairwise disjoint non-empty closed or 
compact subsets $A_0, A_1, \cdots, A_{N-1}$ of the phase space $X$ (here the 
union of all $A_i$'s need not to cover $X$), satisfying certain conditions, 
then the restriction of the system to a invariant subset of $X$ can be 
represented by the full shift on $N$ symbols. When the number of such 
closed (compact) subsets need to be countably infinite, then a subsystem can 
be represented by the full shift with a countable alphabet (\cite{Fu96}).
The family of such closed (compact) subsets with certain conditions is called 
a distillation. More precisely, we give the following definitions. 

\begin{defn}
If for an $f \in M(X)$, there exists an invariant subset $\L \subseteq X$ and 
a subshift $(\S, \s)$ of a certain symbolic dynamical system such that 
$(\S, \s)$ is a symbolic representation of $(\L, f)$, then we call the subshift 
$(\S, \s)$ a {\rm partial representation} of $(X, f)$.
\end{defn}

A partial representation is also helpful to our understanding of the original system, 
especially when $\L$ is a maximal invariant subset or a global attractor of 
$(X, f)$. In these cases, apart from some transient states in 
$X \setminus \L$, all significant dynamical behaviour will asymptotically 
take place in $\L$.  

\begin{defn} \label{def_disti}
Let $X$ be a topological space, $f \in M(X)$. We call a finite family of 
subsets $\MA = \{A_0, A_1, \cdots, A_{N-1} \}$ a {\rm quasi-distillation 
of order N} for the system $(X, f)$  if: \\ 
{\rm (1)} each $A_i$ is compact and non-empty; \\
{\rm (2)} $A_i \cap A_j = \varnothing,~ \forall i \ne j$;\\ 
{\rm (3)} $f(A_i)\supseteq \bigcup _{j \in a(i)}A_j,~ \forall 0\le i \le N-1$.\\
Where $a(i)\subseteq \{0, 1, \cdots, N-1\}$, and each $a(i)$ is non-empty 
and is maximal in the sense that $\forall j \notin a(i), ~ f(A_i)\cap A_j = 
\varnothing$.

We call $\MA$ a {\rm distillation of order N} for $(X, f)$ if it satisfies a 
further condition: \\
{\rm (4)} $\card\bigcap_{s=0}^{+\infty} f^{-s}(A_{i_s})\le 1,~ 
\forall~ (i_0i_1\cdots i_s \cdots) \in \S(N)$.\\
Where $\card(\cdot)$ denotes the cardinal number of a set.
\end{defn}

\begin{defn} \label{def_distil}
Let $X$ and $f$ be as above. We call a countable family of subsets 
$\MA = \{A_0, A_1, \cdots, A_k, \cdots \}$ a {\rm quasi-distillation 
of order infinity} for the system $(X, f)$  if:\\ 
{\rm (1)} each $A_i$ is closed and non-empty;\\
{\rm (2)} there are open subsets $O_i \subseteq X, ~ i \in \Z_+$, such that 
$$
\forall~ i \in \Z_+ , ~ A_i \subseteq O_i  ~~\mbox{and}~~ 
O_i \cap (\bigcup_{i\ne j}A_j)= \varnothing;
$$
{\rm (3)} $f(A_i)\supseteq \bigcup_{j\in a(i)}A_j,~ \forall i \in \Z_+$.\\
Where $a(i)\subseteq \Z_+$, and each $a(i)$ is non-empty 
and is maximal in the same sense as in Definition~\ref{def_disti}.

We call $\MA$ a {\rm distillation of order infinity} for a self-map on a 
metric space $(X, d)$ if it satisfies a further condition:\\
{\rm (4)} $\lim_{n \to + \infty}D(\bigcap_{s=0}^{n} f^{-s}(A_{i_s}))=0,~ 
\forall~ (i_0i_1\cdots i_s \cdots) \in \S(\Z_+)$.\\
Where $D(S)$ denotes the diameter of a subset $S$, i.e., 
$D(S)=\sup_{a,b \in S} d(a, b)$.
\end{defn}

At first we give a general lemma.
\begin{lem} \label{lem_equal}
For a finite or countable family of subsets $\{ A_i, i\in \MI\}$ of a 
topological space $X$ and a map $f: X \to X$, if 
$f(A_j)\supseteq \bigcup _{i \in \MI}A_i,~ \forall j \in \MI$, 
then $\forall l \ge 1, \forall i_k \in \MI, k=0,1, \cdots,$
$$
f^l (\bigcap_{s=0}^{l} f^{-s}(A_{i_s}))= A_{i_l}.
$$
\end{lem}
\proof
{}From 
$$
f(A_j)\supseteq \bigcup _{i \in \MI}A_i,~ \forall j \in \MI,
$$
we have 
$$
f(A_i)\bigcap A_j=A_j, ~ \forall i, j \in \MI.
$$
So when $l=1$, we have
$$
f(A_{i_0}\bigcap f^{-1}(A_{i_1}))\subseteq f(A_{i_0})\bigcap A_{i_1}=A_{i_1}.
$$

$\forall x \in A_{i_1}$, since $f(A_{i_0})\supseteq A_{i_1}, 
\exists y \in A_{i_0}$ such that $f(y)=x$. Therefore $y \in f^{-1}(A_{i_1})$, 
and hence $y \in A_{i_0}\cap f^{-1}(A_{i_1})$, and 
$x=f(y)\in f(A_{i_0}\cap f^{-1}(A_{i_1}))$. That is, 
$$
A_{i_1}\subseteq f(A_{i_0}\bigcap f^{-1}(A_{i_1})).
$$

So we get
$$
f(A_{i_0}\bigcap f^{-1}(A_{i_1}))=A_{i_1}.
$$

By the similar arguement the Lemma can be proven inductively.
\qed

Some known results about distillations and symbolic representations are Smale's horseshoe 
theorem (\cite{Smale}), higher dimensional versions of horseshoes 
(\cite{Wiggins88}) and some generalizations to horseshoe-like invariant sets 
(\cite{Fu96, Zhang}), and etc. Below we give some more general results about 
distillations and representations.

The following theorem gives results on partial representations over a 
finite alphabet.
\begin{thm} \label{th_part_quasi}
Suppose $X$ is a Hausdorff space and $f \in C(X)$, and $(X, f)$ has a 
quasi-distillation of order $N$. Then there exists a subshift of finite type 
$(\S, \s)\le (\S(N), \s)$ such that $(\S, \s)$ is a partial quasi-representation of 
$(X, f)$.

If $(X, f)$ has a distillation of order $N$, then there exists a subshift of 
finite type $(\S, \s)\le (\S(N), \s)$ such that $(\S, \s)$ is a partial 
conjugate representation of $(X, f)$.
\end{thm}

\proof
Let $\S_A=\{x\in \S(N): x=(x_0x_1 \cdots x_k \cdots),~a_{x_kx_{k+1}}=1, 
\forall ~ k \ge 0\}$,
where $A$ is the transition matrix,
$$
A=(a_{ij})_{N\times N}, ~~ 
a_{ij}=\left\{\begin{array}{ll} 1, \;\;\;\; j\in a(i) \\ 
0, \;\;\;\; \mbox{otherwise} \end{array} \right. .
$$

{}From Lemma~\ref{lem_equal}, $\forall l \ge 1$,
$$
f^l (\bigcap_{s=0}^{l} f^{-s}(A_{i_s}))= A_{i_l}, \ \ 
\forall (i_0i_1 \cdots) \in \S_A.
$$
So $\bigcap_{s=0}^{+\infty} f^{-s}(A_{i_s})\ne \varnothing$, since $X$ is 
a Hausdorff space. Let
$$
\L=\bigcap_{s=0}^{+\infty} f^{-s}(\bigcup_{i=0}^{N-1}A_i)
  =\bigcup_{(i_0i_1 \cdots)\in \S_A}\bigcap_{s=0}^{+\infty} f^{-s}(A_{i_s}),
$$
then $\L$ is an invariant compact subset for $f$. Define a coding 
$\f: \L \to \S_A$ as: 
$$
\f(x)= (i_0i_1 \cdots i_k \cdots),~~ 
\forall~ x \in \bigcap_{s=0}^{+\infty}f^{-s}(A_{i_s}).
$$
Then $\f$ is surjective.

$\forall x \in \L$, suppose 
$x \in A_{i_0}, f^s (x)\in A_{i_s}, s=0, 1, \cdots, K$. Then 
$$
\f (x)\in \bigcap_{s=0}^{K}\f f^{-s}(A_{i_s}).
$$ 
{}From the continuity of 
$f^s$,there exists a neighborhood $V_s (x)$ of $x$ such that 
$$
f^s (V_s (x)\cap \L) \subseteq A_{i_s},\ s=0, 1, \cdots, K.
$$ 
Let 
$V(x)=\bigcap_{s=0}^K V_s(x)$, then 
$V(x)\cap\L \subseteq \bigcap_{s=0}^{K}f^{-s}(A_{i_s})$. So 
$\forall y \in V(x)\cap\L$, we have 
$\f (y) \in \bigcap_{s=0}^{K}\f f^{-s}(A_{i_s})$. Thus the first $K$ entries 
of $\f (y)$ and $\f (x)$ agree, therefore 
$$
\r (\f (x), \f (y))\le \frac{1}{2^K}.
$$ 
So $\f$ is continuous. 
The commutativity $\f f|_{\L}=\s|_{\S_A} \f$ is obvious. So $(\S_A, \s)$ is 
a partial quasi-representation of $(X, f)$.

When $(X, f)$ has a distillation of order $N$, then $\f$ is also one-to-one. 
Note that $X$ is a Hausdorff space, $\f$ is therefore a homeomorphism, and 
hence $(\S_A, \s)$ is a partial conjugate representation of $(X, f)$.
\qed

The following two theorems give results on partial representations over 
a countable alphabet.
\begin{thm} \label{th_quasi_distil}
Let $X$ be a sequentially compact $T_1$ space and $f \in C(X)$. 
If $(X, f)$ has a quasi-distillation of order infinity, then there exists 
a countable state Markov subshift $(\S, \s)\le (\S(\Z_+), \s)$ such that 
$(\S, \s)$ is a partial quasi-representation of $(X, f)$.
\end{thm}
\proof
$\forall (s_0, s_1, \cdots) \in \S_A$, where $\S_A$ is defined similarly as in 
the proof of Theorem~\ref{th_part_quasi}, but here $A$ is a matrix of infinite 
order. From 
$$
\bigcap_{s=0}^{+\infty} f^{-s}(A_{i_s})=
\bigcap_{l=0}^{+\infty} \bigcap_{s=0}^{l}f^{-s}(A_{i_s}),
$$
$\bigcap_{s=0}^{+\infty} f^{-s}(A_{i_s})$ is the intersection of a decreasing 
sequence of nonempty closed subsets in a sequentially compact $T_1$ space, so 
it is nonempty. Define a coding $\f: \L \to \S_A$ as:
$$
\f(x)= (i_0i_1 \cdots i_k \cdots),~~ 
\forall~ x \in \L := 
\bigcap_{s=0}^{+\infty} f^{-s}(\bigcup_{i=0}^{+\infty}A_i)
  =\bigcup_{(i_0i_1 \cdots)\in \S_A}\bigcap_{s=0}^{+\infty} f^{-s}(A_{i_s}),
$$
then $\f$ is onto. Let $N>0$ satisfy $\ep > \sum_{n=N}^{+\infty}\frac{1}{2^n}$. 
Suppose $x \in A_{i_0}$ and $f^s(x) \in f^s (x)\in A_{i_s}$, 
then $ x\in f^{-s}(A_{i_s})$ and $ \f (x)\in \f f^{-s}(A_{i_s}), 
s=0, 1, \cdots, N$. From the continuity of $f^s$, there exists a neighborhood 
$V_s (x)$ of $x$ such that 
$f^s (V_s (x)) \subseteq O_{i_s},\ s=0, 1, \cdots, N.$  Let
$V(x)=\bigcap_{s=0}^N V_s(x)$, then $f^s (V(x))\subseteq O_{i_s}$, therefore 
$f^s (V (x)\cap \L) \subseteq A_{i_s},\ s=0, 1, \cdots, N,$ and 
$V (x)\cap \L \subseteq \bigcap_{s=0}^N f^{-s}(A_{i_s})$. So 
$y \in V(x)\cap\L$ implies 
$\f (y)\in \f \bigcap_{s=0}^N f^{-s}(A_{i_s})
\subseteq \bigcap_{s=0}^N \f f^{-s}(A_{i_s})$. 
Thus the first $N$ entries of $\f (y)$ and $\f (x)$ agree, hence 
$\r (\f (x), \f (y)) < \ep$. This proves the continuity of $\f$. Therefore 
$(\S_A, \s)$ is a partial quasi-representation of $(X, f)$.
\qed

\begin{thm}
Let $(X, d)$ be a complete metric space and $f \in C(X)$.
If $(X, f)$ has a distillation of order infinity, then there exists a 
countable state Markov subshift $(\S, \s)\le (\S(\Z_+), \s)$ such that 
$(\S, \s)$ is a partial conjugate representation of $(X, f)$.
\end{thm}
\proof
As being shown in the proofs of 
Theorems~\ref{th_part_quasi}~and~\ref{th_quasi_distil},
$\bigcap_{s=0}^{+\infty} f^{-s}(A_{i_s})$ is a nonempty set for all 
$(i_0i_1 \cdots i_k \cdots)\in \S_A$. From the condition (4) in 
Definition~\ref{def_distil}, $\bigcap_{s=0}^{+\infty} f^{-s}(A_{i_s})$ is a 
one-point set. Define a coding $\f: \L \to \S_A$ as:
$$
\f (\bigcap_{s=0}^{+\infty} f^{-s}(A_{i_s}))=(i_0i_1 \cdots i_s \cdots),
$$
where 
$\L=\bigcap_{s=0}^{+\infty} f^{-s}(\bigcup_{i=0}^{+\infty}A_i)
  =\bigcup_{(i_0i_1 \cdots)\in \S_A}\bigcap_{s=0}^{+\infty} f^{-s}(A_{i_s}).$
Then as shown in the proof of Theorem~\ref{th_quasi_distil}, $\f$ is 
continuous.

For $x=(x_0x_1 \cdots) \in \S_A, \ \ \forall \ \ep >0, \ \ \exists \ N >0, $ 
whenever $n \ge N$, we have 
$$
D(\bigcap_{s=0}^{n}f^{-s}(A_{a_s})) < \ep.
$$
$\forall \ y \in \S_A,$ when $\r (x, y)< 1/2^N$, the first $N+1$ entries of 
$x$ and $y$ agree. Thus
$$
d(\f ^{-1}(x), \f ^{-1}(y))\le D(\bigcap_{s=0}^{N}f^{-s}(A_{x_s})) < \ep,
$$
hence $\f ^{-1}$ is continuous. This shows that $\f$ is a homeomorphism. 
Therefore $(\S_A, \s)$ is a partial  conjugate representation of $(X, f)$.
\qed

\section{Representations for Discontinuous Maps}  \label{discontinuous}

This section will discuss  some specific
examples and show that it is possible to use symbolic dynamics as a tool for 
further studies of dynamics of a class of discontinuous maps: piecewise 
continuous maps.


Let $X \subseteq \R^n$, and $\MP= \{ P_0, P_1, \cdots, P_{N-1}\}$ be 
a finite family of subsets of $X$, satisfying $\bigcup_{i=0}^{N-1}P_i = X$, 
and $P_i \cap P_j=\varnothing$ for $i\ne j$. A piecewise continuous map 
is a map $f: X \to X$ whose restriction to each 
 $P_i,~ 0\le i \le N-1$, is continuous, and $\MP$ is minimal in the sense 
that $f$ is not continuous on $P_i \cup P_j$ for $i \ne j$.

A partition $\MP = \{ P_0, \cdots, P_{N-1}\}$ associated to a piecewise 
continuous map provides a natural coding $\f: X \to \S(N)$ by 
$\f(x) =  (i_0 i_1 \cdots i_k \cdots)$ where $f^k(x)\in P_{i_k}$.

Let $G=\{ (x, \f(x)),~ x \in X\}$ be the graph of $\f: X \to \S(N)$ endowed 
with the metric
$$
d_G((x, \f(x)), (y, \f(y))) = \max \{ d(x,y), \r(\f(x), \f(y))\},
$$
where $d$ is the metric on $X$. Then the extension map 
$f_G: G \to G, ~ f_G(x, \f(x))=(f(x), \f(f(x)))$, is continuous (\cite{Goetz}). 
This result may be useful since sometimes it is a bridge between discontinuous 
and continuous maps.

Here we give a definition of partitions for piecewise continuous maps.

\begin{defn}
For $X \subseteq \R^n,~f \in M(X)$. We call a finite family of subsets 
$\MP = \{ P_0, P_1, \cdots, P_{N-1}\}$ a {\rm partition} of $(X, f)$ if: 
{\rm (1)} each $P_i$ is open and convex;
{\rm (2)} $P_i \cap P_j = \varnothing$ for $i \ne j$;
{\rm (3)} $X= \bigcup_{i=0}^{N-1}\overline{P_i}$;
{\rm (4)} $\forall ~ 0\le i \le N-1,~ f|_{P_i}$ is continuous and can be 
          extended to a continuous map on $\overline{P_i}$.

We call a partition {\rm minimal} if $f$ is not continuous on 
$\overline{P_i} \cup \overline{P_j}$ for $i \ne j$.

A {\rm cell}, denoted by $C(x)$, is a set of points in $X$ encoded by the 
same symbolic sequence $\f(x)$, i.e., $C(x)= \{ y\in X: \f(y)=\f(x)\}$, 
where $\f$ is called the {\rm coding} associated with the partition 
$\MP$, $\f: X \to \S(N)$ is defined by 
$\f(x) =  (i_0i_1\cdots i_k\cdots)$ where $f^k(x)\in P_{i_k}$.
\end{defn}

\begin{eg} \label{eg_gauss}
Symbolic representation for the Gauss map.
\end{eg}

Let $ ( I , g )$ be the iterated system of the Gauss map (\cite{Adler91})
$ g : x \rightarrow x^{-1} ~(\bmod ~1)$ for $0< x \le 1$, and $g(0)=0$.
Let a sequence
$s = ( s_{n})_{n \in \Z_+} \in \S(\N)$  correspond to the continued fraction
 expansion
$[s_{0}s_{1}s_{2}\cdots]$.
This defines the map $\pi $ from
$\S(\N)$ to $I=[0, 1]$:
$$
\pi ( s_{0} \,, s_{1} \,, \cdots )=[s_{0}s_{1}s_{2}\cdots].
$$
It can be verified that
\begin{itemize}
\item[{(i)}] $g \pi = \pi \s $,
\item[{(ii)}] $\pi $ is continuous,
\item[{(iii)}] $\pi $ is onto,
\item[{(iv)}] there is a bound on the number of pre-images (in this
case two),
\end{itemize}and
\begin{itemize}
\item[{(v)}] there is a unique pre-image of ``most" numbers in $I$ (here those
with infinite continued fraction expansions).
\end{itemize}
So $(\S(\N), \s)$ is a regular representation of $(I, g)$. 
The map $\pi $ is not a homeomorphism, but we do have
a satisfactory representation of the dynamical system
by a one-sided full shift over a countable alphabet in the sense 
of Adler \cite{Adler}, i.e., 
\begin{itemize}
\item orbits are preserved;
\item every point has at least one symbolic representative;
\item there is a finite upper limit to the number of representatives
      of any point; 
\item and every symbolic sequence represents some point.
\end{itemize}

There is an alternate definition of $\pi $ in terms of
a countable partition of $I$:
$$ 
\MR = \{ R_{i}, i \in \N \}, \ \ R_i = (\frac{1}{i+1}, \frac{1}{i}).
$$
$$
\pi ( s_{0} , s_{1} , \cdots ) = \bigcap _{ n = 
0}^{\infty }\overline{ R_{s_{0}} \cap g^{-1} ( R_{s_{1}}) \cap \cdots 
\cap g^{-n} (R_{s_{n}})}.
$$

Note that in a suitable torus topology the Gauss map becomes continuous 
everywhere except at $x=0$. Below we give another example where the map is 
``more discontinuous'' but is still easy to symbolize.

\begin{eg} \label{eg_interval} 
Symbolic representations for interval exchange transformations.
\end{eg}

An interval exchange transformation $f: I \to I, I=[0, 1]$, over a partition 
$\MP=\{I_0, I_1, \cdots, I_{N-1}\}$ is a one-dimensional piecewise isometry, 
so we can follow the arguements in \cite{Ashwinf, Ashwinft, Goetz00} to discuss 
symbolic representation for $f$.

By naturally coding orbits of $f$ using the partition $\MP, \f: I \to \S(N)$,
$$
\f(x)=(\cdots i_{-1} i_0 i_1 \cdots),
$$ 
where $f^k(x)\in I_{i_k}, k \in \Z$, we obtain a subset $\S \subseteq \S(N)$ of 
bi-infinite symbolic sequences, $\S=\{ \f(x)\in \S(N), x \in I\}$.
$\S$ is closed and  shift invariant, so $(\S, \s)\le (\S(N), \s)$.

Note that if a subinterval $S \subseteq I$ is invariant under $f^m$, then 
$f^m |_S$ is either the identity or the reflection in the midpoint of $S$ (in this 
case $f^{2m} |_S$ is the identity). So the disk/polygon packing discussed in 
\cite{Ashwinf, Ashwinft, Goetz00} now reduce to rigid interval (\cite{Katok}) 
packing, i.e., the set
$$
J=\overline{I\setminus \bigcap_{n \in \Z}f^n (\hat{I})},~ 
\hat{I}= \bigcup_{k=0}^{N-1} {\rm int} I_k
$$
of points that never hit or approach the discontinuity can be decomposed into 
finite (\cite{Katok}:Lemma 14.5.4) cells, or rigid intervals $J_k$,
$$
J=\bigcup_{k=1}^K J_k,~ K\le 2(N-1),
$$ 
all points in $J$ have periodic codings, and $f$ is periodic on each $J_k$.

As discussed in \cite{Katok}, when $f$ is generic (i.e., $J=\varnothing$), we 
can define a map $h: \S \to I$,
$$
h((\cdots i_{-1} i_0 i_1 \cdots))=\bigcap_{k\in \Z}f^{-k}(I_{i_k}).
$$
$h$ satisfies:
\begin{itemize}
\item[{(1)}] $f h = h \s $,
\item[{(2)}] $h $ is continuous,
\item[{(3)}] $h $ is onto,
\item[{(4)}] there is a bound on the number of pre-images (in this
case $2^N$),
\end{itemize}and
\begin{itemize}
\item[{(5)}] there is a unique pre-image of ``most" points in $I$ (here those
no image or pre-image of them are discontinuity points).
\end{itemize}
So $(\S, \s)$ is a regular representation of $(I, f)$. 
Although the map $h$ is not a homeomorphism, we do have
a satisfactory representation of the dynamical system
by a two-sided subshift over a finite alphabet with the properties listed 
near the end of Example~\ref{eg_gauss}.

If $(\S, \s)$ is topologically transitive, then $(\S, \s)$ is a subshift 
of {\em infinite} type, because otherwise the periodic points of $\s|_{\S}$ 
would be dense in $\S$.

%
%

\section{Final Remarks}

When we consider the symbolic representation of an iterated map, 
one may hope to find one of the following:
\begin{itemize}
\item[{(A)}] a full representation (i.e., all points in the phase space 
are symbolically represented);
\item[{(B)}] a conjugate representation;
\item[{(C)}] a representation via a subshift of finite type over a finite 
or countable alphabet.
\end{itemize}
Unfortunately, it is very rare to get a symbolic representation satisfying all 
of (A), (B) and (C). Usually, if we'd like to get a symbolic representation 
satisfying (A) and (B), we may need to pay the price of giving up (C), i.e., 
we need to use an uncountable alphabet, as discussed in 
Sections~\ref{conjugate} and \ref{sec_second}; if we'd like to obtain a 
representation with properties (A) and (C), we may have to give up (B), 
as discussed in Sections~\ref{quasi} and \ref{partition}; if we'd like to 
find a representation with (B) and (C), we may have to give up (A), i.e., 
we only get a partial representation, as discussed in 
Section~\ref{distillation}.

When we consider a symbolic representation of a discontinuous map, we at least 
have to give up (B), as discussed in Section~\ref{discontinuous} using special 
cases of piecewise continuous maps.

Symbolic representations provide a powerful method to investigate 
general discrete dynamical systems through shifts or subshifts. This is 
similar to the situation for the study of finite groups in algebra, where  
all finite groups can be represented by subgroups of symmetric groups $S_n$,
which are better understood.  We can view $S_n$ as `symbolic' groups. 

There is a large body of theory that describes the dynamics of
continuous smooth dynamical systems. However, if there are discontinuities 
in the system such as those caused by collisions (impacting) or
switching there is still comparatively little in the way of general theory to
describe such systems. In Section~\ref{discontinuous} we have shown through 
specific examples that it is helpful to use symbolic dynamics to develop 
the general theory of dynamics of a class of discontinuous maps: piecewise 
continuous maps. 
Most results in this paper are proven constructively. This makes our 
results potentially useful for applications.

Finally, we conjecture that some results in 
Sections~\ref{conjugate}, \ref{sec_second}, \ref{quasi} and \ref{discontinuous} 
about representations for one 
dimensional maps may be extended to higher dimensional cases. 
Higher dimensional binary expansions and higher dimensional continued 
fractions (\cite{Arnold}) may be useful in the extension. We hope to discuss 
this topic in a separate paper.

\vspace{0.5cm}

\noindent {\bf Acknowledgments.} 
{\small X. Fu and P. Ashwin thank the EPSRC for 
 support via grant GR/M36335, and they are grateful to Surrey University 
 where earlier versions of this paper were written. The authors would 
 like to thank Prof. Robert Harrison for his helpful comments and advice.
 While an earlier version of this paper was written,
 X. Fu was supported jointly by a scholarship from the Nonlinear 
 Dynamics Group in Physics Department of Heriot-Watt University and 
 a grant (No. 19572075) from China National Natural Science Foundation.}

\end{document}